\documentclass[%
reprint,
superscriptaddress,
 amsmath,amssymb,
 aps,
]{revtex4-2}

\usepackage{graphicx}
\usepackage{dcolumn}
\usepackage{hyperref}
\usepackage{xcolor}

\begin{document}

\preprint{APS/123-QED}

\title{Structural changes across thermodynamic maxima in supercooled liquid tellurium:\\
a water-like scenario}

\author{Peihao Sun}
\email[Corresponding author: ]{peihao.sun@unipd.it}
\affiliation{SLAC National Accelerator Laboratory, 2575 Sand Hill Rd, Menlo  Park, CA 94025, USA}
\affiliation{Dipartimento di Fisica e Astronomia ``Galileo Galilei'', Universit{\`a} degli Studi di Padova, Via F. Marzolo, 8, 35131 Padova, Italy}

\author{Giulio Monaco}
\affiliation{Dipartimento di Fisica e Astronomia ``Galileo Galilei'', Universit{\`a} degli Studi di Padova, Via F. Marzolo, 8, 35131 Padova, Italy}

\author{Peter Zalden}
\affiliation{European XFEL, Holzkoppel 4, 22869 Schenefeld, Germany}

\author{Klaus Sokolowski-Tinten}
\affiliation{Faculty of Physics and Center for Nanointegration Duisburg-Essen (CENIDE), University of Duisburg-Essen, Lotharstrasse 1, 47048 Duisburg, Germany}

\author{Jerzy Antonowicz}
\affiliation{Faculty of Physics, Warsaw University of Technology, Koszykowa 75, Warsaw, 00-662, Poland.}

\author{Ryszard Sobierajski}
\affiliation{Institute of Physics of the Polish Academy of Sciences, PL-02-668 Warsaw, Poland}

\author{Yukio Kajihara}
\affiliation{Graduate School of Advanced Science and Engineering, Hiroshima University, Higashi-Hiroshima, Hiroshima 739-8521, Japan}

\author{Alfred Q. R. Baron}
\affiliation{Materials Dynamics Laboratory, RIKEN SPring-8 Center, 1-1-1 Kouto, Sayo, Hyogo 679-5148, Japan}

\author{Paul Fuoss}
\affiliation{SLAC National Accelerator Laboratory, 2575 Sand Hill Rd, Menlo  Park, CA 94025, USA}

\author{Andrew Chihpin Chuang}
\author{Jun-Sang Park} 
\author{Jonathan Almer}
\affiliation{X-ray Science Division, Advanced Photon Source, Argonne National Laboratory, Lemont, Illinois 60439, USA}

\author{J. B. Hastings}
\email[Corresponding author: ]{jbh@slac.stanford.edu}
\affiliation{SLAC National Accelerator Laboratory, 2575 Sand Hill Rd, Menlo  Park, CA 94025, USA}

\date{\today}

\begin{abstract}
  Liquid polymorphism is an intriguing phenomenon which has been found in a few single-component systems, the most famous being water. By supercooling liquid Te to more than 130\,K below its melting point and performing simultaneous small-angle and wide-angle X-ray scattering measurements, we observe clear maxima in its thermodynamic response functions around 615\,K, suggesting the possible existence of liquid polymorphism. A close look at the underlying structural evolution shows the development of intermediate-range order upon cooling, most strongly around the thermodynamic maxima, which we attribute to bond-orientational ordering. The striking similarities between our results and those of water, despite the lack of hydrogen-bonding and tetrahedrality in tellurium, indicate that water-like anomalies may be a general phenomenon among liquid systems with competing bond- and density-ordering.
\end{abstract}

\maketitle

\section{Introduction}
Liquid polymorphism, or the existence of two or more liquid phases of the same substance, is a topic of much current interest~\cite{Stanley2013,Tanaka2020}. Although the concept dates back to at least the 1960s~\cite{Rapoport1967}, supporting evidence had been lacking until recent years, when observations of phenomena related to liquid polymorphism began to emerge in a variety of systems. Perhaps the most intriguing among these are single-component systems, where the polymorphism points out the insufficiency of the theory of simple liquids~\cite{Hansen2006} to describe real, albeit compositionally simple, fluid systems, while suggesting the strong influence of local structures  even for the properties of amorphous materials. The current list of systems with clear experimental evidence for a liquid-liquid transition (LLT), in decreasing complexity, includes large molecules such as triphenyl phosphite~\cite{Kurita2004} and atomic systems such as phosphorus~\cite{Katayama2000,Monaco2003}, sulfur~\cite{Henry2020}, and cerium~\cite{Cadien2013}. Meanwhile, an LLT has been suggested in many other systems with varying degrees of experimental and theoretical evidence~\cite{Tanaka2020}. 

Among these systems, water is perhaps the most famous. It is well-known that water exhibits a number of thermodynamic anomalies such as a density maximum at 4 $^\circ$C under ambient pressure. Among the different scenarios explaining water's anomalies~\cite{Pallares2014,Gallo2016}, the second critical point hypothesis~\cite{Poole1992} has emerged to be a leading theory, 
as it is observed in recent simulations of water models~\cite{Debenedetti2020,Palmer2018} while also consistent with experimental results~\cite{Kim2017,Kim2020a} (although the latter ones are not without controversy~\cite{Caupin2018,Kim2018}).
Under this framework, there exists an LLT in water in the deeply supercooled region and under elevated pressures, and the coexistence line of this LLT terminates at a liquid-liquid critical point (LLCP). A peculiar aspect of the LLT in liquid water, distinguishing it from most other systems, is the fact that the low-density liquid phase lies on the low temperature side of the transition. While this explains the density maximum in a straightforward way~\cite{Poole1992,Gallo2016}, it also leads to the distinctive phase diagram where the LLT exists at higher pressures than the LLCP, thus setting water apart from other single-component systems mentioned above. In addition, the critical pressure is predicted to be positive but not far from zero, so thermodynamic effects related to critical fluctuations can be observed along the $P \approx 0$ isobar~\cite{Kim2017}. These effects manifest as maxima in thermodynamic response functions such as the isothermal compressibility, and they form a line emanating from the LLCP usually referred to as the Widom line~\cite{Xu2005}. For ordinary water (H$_2$O), evidence suggests that the Widom line intersects the $P \approx 0$ isobar around 229 K~\cite{Kim2017}. A schematic phase diagram proposed for water is shown in Fig.~S1.

In general, the physical origin of water's peculiarities is not entirely clear, and it remains an open question whether they are unique to water or are general to a family of liquids. In fact, it has been suspected that other tetrahedrally coordinated systems (such as silicon) exhibit similar behaviors~\cite{Vasisht2011,Beye2010}. Another candidate is liquid tellurium (Te), whose lack of tetrahedrality makes it a somewhat surprising example. As observed first by Kanno et al.~\cite{Kanno2001} and reiterated by Angell~\cite{Angell2007a}, liquid Te also exhibits a variety of thermodynamic anomalies which bear a striking resemblance to those of water. A systematic study of the properties of liquid Te thus can be informative on the universality of water-type LLT, bearing significance from a fundamental physics aspect. From a practical point of view, Te is the basis element of several phase-change materials (PCMs) that are becoming increasingly important for information technology~\cite{Wong2010}. Thus, it is reasonable to expect that the phenomena attributed to liquid polymorphism in Te-based PCMs~\cite{Lucas2020} are related to a possible LLT in liquid Te.

Therefore, in this work, we study the properties of bulk supercooled liquid Te under low vapor pressure ($< 0.5$ mbar). We observe maxima in various thermodynamic response functions, including the isothermal compressibility, around 615 K. Moreover, we are able to determine details of the structural changes across the thermodynamic response maxima, which show many similarities with water. Our results clearly show that intermediate-range ordering plays a central role in the structural modification, and indicate that water-like thermodynamic anomalies may be attributed to a competition between density- and bond-ordering regardless of the details of the local structure.

\section{Simultaneous SAXS and WAXS}
In order to systematically study the thermodynamic and structural changes, we performed simultaneous small-angle and wide-angle X-ray scattering (SAXS and WAXS) measurements at beamline 1-ID at the Advanced Photon Source. The high X-ray photon energy, 76.112 keV, allows us to probe bulk Te samples in a simple transmission geometry (see Materials and Methods for details).

Our dataset includes three scans. The scans are temporally separated during the experiment, performed on different samples, and the X-ray beam condition differs between the scans. The consistency of the results from all scans, as will be shown below, strongly supports our conclusions. Each scan consists of several rounds of temperature cycling, where in each round the sample is first heated above its melting point of 723 K and then cooled down in steps (usually 10 K) to the lowest temperature. At each step, after temperature equilibration, X-ray scattering patterns are collected for several minutes. Recrystallization happens suddenly and can be identified by the appearance of sharp Bragg peaks in the WAXS pattern; these data are excluded from the analysis.

\begin{figure}
    \centering
    \includegraphics[width=0.95\columnwidth]{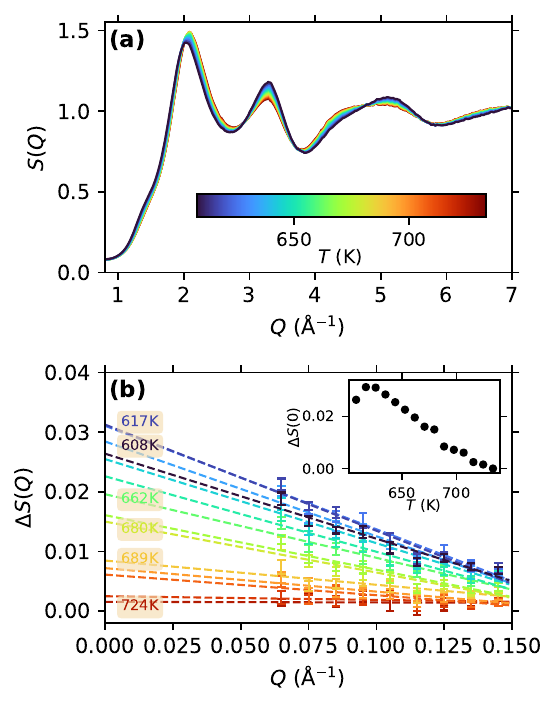}
    \caption{(a) Te WAXS profiles (structure factor $S(Q)$) from scan 1. From dark red to dark blue are decreasing temperatures as indicated by the color bar. (b) Te SAXS profiles from the same scan, subtracting the highest temperature profile (733 K) in the scan. The color scheme is the same as in (a). Data are shown with error bars, and the dashed lines show linear fits to the data. Several temperatures are marked next to the fit lines with the corresponding colors. Inset shows the intercept $\Delta S(0)$ plotted against sample temperature; the error bars are smaller than the symbol.}
    \label{fig:Te_raw_scan1}
\end{figure}

Figure~\ref{fig:Te_raw_scan1} shows the WAXS and SAXS results from scan~1. The structure factor $S(Q)$ is obtained from the WAXS patterns, and the results agree well with those in the literature (see SI Appendix). Several trends can be clearly observed. Firstly, with cooling, the first peak moves to lower $Q$ values, which is similar to the behavior of water~\cite{Kim2017,Benmore2019}. The second peak becomes sharper while its position remains relatively constant at 3.3 \AA$^{-1}$. The broad third peak spanning 4 to 6 \AA$^{-1}$ gradually changes its symmetry upon cooling, with the low-$Q$ side centered around 4.4 \AA$^{-1}$ decreasing in magnitude and the high-$Q$ side centered around 5.1 \AA$^{-1}$ growing. Remarkably, the same phenomenon has been observed in \emph{ab initio} molecular dynamics (AIMD) simulations~\cite{Akola2010}, although the growth of the high-$Q$ side becomes obvious in that study only at the very low temperature of 560 K.

An interesting feature is the presence of a pre-peak centered around 1.4 \AA$^{-1}$.
More detailed analysis of this pre-peak will be presented later. We note here that a pre-peak is also observed in Te-based chalcogenides such as Ge$_{15}$Te$_{85}$~\cite{Bergman2003,Kalikka2012,Wei2017,Gaspard2016}, as well as phase-change materials~\cite{Zalden2019} particularly in the supercooled liquid region. In these materials, the appearance of the pre-peak is attributed mostly to the Ge atoms~\cite{Kalikka2012} and explained as a Peierls-like distortion~\cite{Wei2017,Gaspard2016}. Here in pure Te, the pre-peak is found to exist above the melting point as well, where the liquid is metallic~\cite{Cabane1971}. Therefore, a Peierls-like distortion is unlikely, while it is possible that the pre-peak arises from bond-orientational order, as will be discussed later.

For SAXS, as in most other atomic liquids, the structure factor of liquid Te is much less than unity and the scattering is weak. Consequently, the background is significant and its pattern depends on the position of the sample and the furnace, and thus it is difficult to be measured and subtracted accurately. Nonetheless, within each scan, sample movement is negligible and the background pattern is expected to remain unchanged. Therefore, for the SAXS analysis we calculate the change in the structure factor, $\Delta S(Q)$, from the highest temperature in each scan, which cancels out the background. For scan 1, this is 733 K, above the melting point at 723 K. The results are shown in Fig.~\ref{fig:Te_raw_scan1}b, and they show unambiguously a non-monotonic behavior with cooling: while $\Delta S(Q)$ increases at first, it begins to decrease at the lowest temperatures. To quantify the change, as in earlier works on chalcogen systems~\cite{Kajihara2012}, we fit a linear function to $\Delta S(Q)$ with the results shown as dashed lines. The extracted intercept, $\Delta S(0)$, is plotted against temperature in the inset, and a peak can be clearly observed. This suggests the existence of thermodynamic maxima to be discussed below.

\section{Maxima in the thermodynamic response functions}
\begin{figure}
    \centering
    \includegraphics[width=0.8\columnwidth]{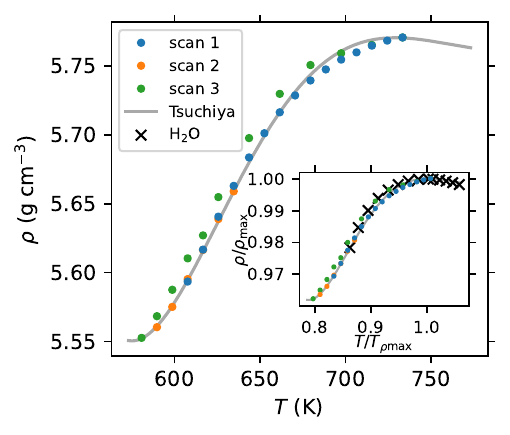}
    \caption{Density of liquid Te plotted against temperature. Dots with different colors show results from the three different scans, which agree well with each other and with the data reported by Tsuchiya~\cite{Tsuchiya1991} (gray line). Inset shows the same data in reduced temperature and density, which are normalized by the location of density maximum $\rho_\mathrm{max}=5.77$ g/cm$^3$ around $T_{\rho\mathrm{max}}=729$ K. Black crosses show the results on supercooled~\cite{Speedy1987} and normal liquid water~\cite{Wagner2002}, which bear remarkable resemblance to Te.}
    \label{fig:Te_density}
\end{figure}

The non-monotonic behavior in the SAXS region bears much resemblance to recent results on supercooled water~\cite{Kim2017}, where it is found to associate with the liquid-liquid transition. We now look further into the extent of similarities between the two systems. The density maximum at 4 $^\circ$C is often thought to be a salient anomaly of water. Figure~\ref{fig:Te_density} shows our results from all three scans on the density of supercooled liquid Te obtained with X-ray transmission measurements (see SI Appendix). The results are consistent between scans, and all agree well with the values previously reported by Tsuchiya~\cite{Tsuchiya1991}. The inset shows the same data plotted on reduced scales normalized by the maximum density $\rho_\mathrm{max}=5.77$ g/cm$^3$ and the corresponding temperature $T_{\rho\mathrm{max}}=729$ K. When the available data on water is plotted on the same reduced scale (with $\rho_\mathrm{max}=1$ g/cm$^3$ and $T_{\rho\mathrm{max}}=277$ K), a remarkable resemblance between the two systems can be observed, as has been pointed out in the literature~\cite{Kanno2001,Angell2007a}.
In comparison, other tetrahedral systems such as silica (in experiment)~\cite{Angell1976} and silicon (in simulation)~\cite{Angell1996} show a density maximum that is significantly broader. Thus, the similarities between Te and water may suggest that the degree of structural ordering is comparable between them~\cite{Shi2018}. 

\begin{figure}
    \centering
    \includegraphics[width=0.95\columnwidth]{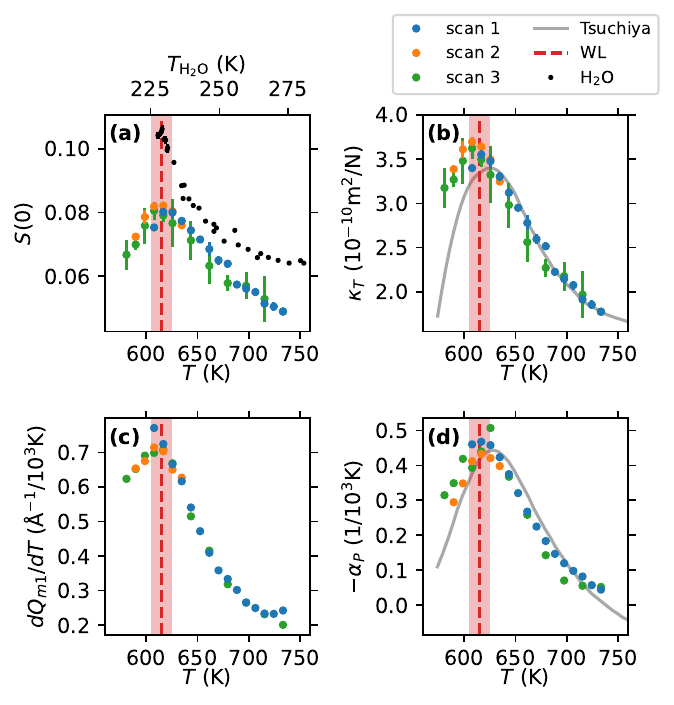}
    \caption{Maxima in the properties of supercooled Te. (a) $S(0)$, (b) isothermal compressibility, $\kappa_T$, (c) temperature derivative of the first peak position, $Q_{m1}$, and (d) the opposite of the thermal expansion coefficient $\alpha_P$, all plotted against sample temperature $T$. Results by Tsuchiya~\cite{Tsuchiya1991} are shown as gray lines in (b) and (d). The vertical dashed line shows the approximate position of the maxima, 615 K, and the red shaded region indicate a $\pm 10$\,K range around it as a guide to the eye. In (a) we also plot $S(0)$ for water (small black dots, upper $x$-axis)~\cite{Kim2017}.}
    \label{fig:Te_SAXS_summary}
\end{figure}

As in the case of water~\cite{Gallo2016}, the density anomaly in liquid Te hints at the possible existence of liquid polymorphism. To provide further evidence for this scenario, in Fig.~\ref{fig:Te_SAXS_summary} we show, with colored symbols representing different scans, four properties of liquid Te: (a) the $Q \to 0$ limit of the structure factor, $S(0)$, (b) the isothermal compressibility, $\kappa_T\equiv -V^{-1}(\partial V/\partial P)_T$, (c) the temperature derivative of the position of the first peak, $Q_{m1}$, and (d) the opposite of the thermal expansion coefficient $\alpha_P \equiv V^{-1}(\partial V/\partial T)_P$. Of these quantities, $Q_{m1}$ can be obtained from the WAXS data in a straightforward way, and $\alpha_P$ can be derived from the density measurement shown above. To obtain $S(0)$ and $\kappa_T$, we use the following relation~\cite{Svergun1987}:
\begin{equation} \label{eq:S0_kT_relation}
    S(0) = k_B T \frac{\rho}{m} \kappa_T, 
\end{equation}
where $k_B$ is the Boltzmann constant and $m$ is the mass of the Te atom. For $S(0)$, the value at the highest temperature for each scan is determined using Eq.~(\ref{eq:S0_kT_relation}) and a known value of $\kappa_T$: for scans 1 and 3, this is above the melting point and we use the value given in Ref.~\cite{Tsuchiya1991}; for scan 2, we use the results from the other two scans. Because we have measured the relative change $\Delta S(0)$ from the highest temperature, the values of $S(0)$ and $\kappa_T$ are then obtained.
The presence of maxima is evident in all panels of Fig.~\ref{fig:Te_SAXS_summary}. For $S(0)$ and $\kappa_T$ we note that, although the procedure described above is used to obtain their absolute values, the maxima can already be seen in $\Delta S(0)$ as shown in Fig.~\ref{fig:Te_raw_scan1}b as well as Figs.~S10b and S11b in SI Appendix. The approximate position of the maxima, 615 K, is marked with a dashed vertical line in Fig.~\ref{fig:Te_SAXS_summary}, while the shaded region indicates a $\pm 10$ K range as a guide to the eye.

The existence and the shape of these maxima provide crucial information on the thermodynamics of liquid Te, in particular the possible morphology of its liquid phase diagram, which has not been discussed much in the literature. Given the similarities between the thermodynamic properties of Te and water, however, it is reasonable to refer to studies on the latter, of which an extensive amount exist. Specifically, to explain water's thermodynamic anomalies most of which are shared by Te, four scenarios have been proposed~\cite{Pallares2014,Gallo2016}: (i) the ``retracting spinodal'' scenario~\cite{Speedy1982}, where the anomalies are consequences of the liquid-vapor spinodal re-entering the supercooled liquid region; (ii) the ``second critical point'' scenario~\cite{Poole1992}, where the anomalies are effects related to an LLCP located in the supercooled region at positive pressures; (iii) the ``hidden critical point'' scenario~\cite{Poole1994}, where the LLCP is located in the unstable region defined by the liquid-vapor spinodal and is thus unreachable; (iv) the ``singularity free'' scenario, where the thermodynamic maxima exist but there is no critical point on the phase diagram~\cite{Sastry1996}. For liquid Te, the maxima observed in Fig.~\ref{fig:Te_SAXS_summary} rule out scenario (i) which predicts divergence due to the spinodal~\cite{Speedy1982}, while the breadth of these maxima and the continuous change in the structure factor (e.g., $Q_{m1}$) excludes scenario (iii) where a first-order LLT is expected~\cite{Poole1994}. 

Therefore, we are left with the LLCP scenario (ii) and the singularity-free scenario (iv). The present results do not fully confirm or exclude either case, since the fluctuations observed here are not large enough to be attributed to large-scale critical fluctuations, which only exist very close to a critical point. However, it should be noted that the mechanism underlying the singularity-free models proposed so far is the existence of two local states in the liquid and the interconversion between them, which is often referred to as the ``two-state model''~\cite{Gallo2016,Caupin2021,Tanaka2012,Tanaka2020}. Several of these models can be understood as variations of the LLCP model where the LLCP is inaccessible, for example because it is located at $T=0$ K~\cite{Sastry1996,Stokely2010} or because the critical density is too high to be reached~\cite{Caupin2021}. In those cases, the physics governing the behavior of the liquid is the same as that for the ``supercritical'' states in the LLCP scenario, which include the $P=0$ isobar. Nonetheless, given the lack of an actual LLCP in those cases, it may not be accurate to use the term ``Widom line'' to describe the thermodynamic response maxima; instead, the term ``Schottky line'' derived under the framework of the two-state model has been used in some works~\cite{Tanaka2020,Shi2020a}. Since our results do not distinguish between the two scenarios, in this work we simply use the phrase ``thermodynamic response maxima'' to refer to their approximate position, 615 K. It should be mentioned that similar controversies surround the interpretation of results on supercooled water as well~\cite{Tanaka2020,Caupin2021}; perhaps the most convincing proof would be to find the LLCP with critical-like behavior in its vicinity, as has been done in the molecular dynamics simulations of several water models~\cite{Debenedetti2020,Tanaka2020}.

Notably, the maxima in all quantities shown in Fig.~\ref{fig:Te_SAXS_summary} are close to each other in temperature. This is an important observation because different quantities are expected to have a different background contribution given by the behavior of a normal liquid~\cite{Shi2020a}. In particular, for ordinary liquids, $\kappa_T$ increases with temperature, so the peak in $\kappa_T$ is expected to shift to somewhat higher $T$; on the other hand, $-\alpha_P$ decreases with temperature for normal liquids, so the peak in $-\alpha_P$ should be shifted to lower $T$. The fact that the peaks in Fig.~\ref{fig:Te_SAXS_summary}b and d are very close in temperature means that the shifts are small; this is also consistent with the sharpness of the peaks. Therefore, 615 K should be a good approximation of the position of the Schottky or Widom line for liquid Te at $P \approx 0$.

Further evidence for the similarity between Te and water can be seen in Fig.~\ref{fig:Te_SAXS_summary}a, where the measured values of $S(0)$ in supercooled water~\cite{Kim2017} are also shown. Note that $S(0)$ is a dimensionless quantity reflecting the number fluctuations in the system~\cite{Svergun1987}:
\begin{equation}
    S(0) = \frac{\langle (N- \langle N \rangle)^2 \rangle}{\langle N \rangle},
\end{equation}
where $N$ is the number of particles (Te atoms or water molecules) in a given sample volume and the angular brackets denote the ensemble average. Thus, the fact that $S(0)$ is close in magnitude between Te and water at $P \approx 0$ indicates that, if the LLCP exists, the critical pressure in liquid Te may be as close to zero pressure as in water.

\section{Structural changes}
\begin{figure*}
    \centering
    \includegraphics[width=0.7\textwidth]{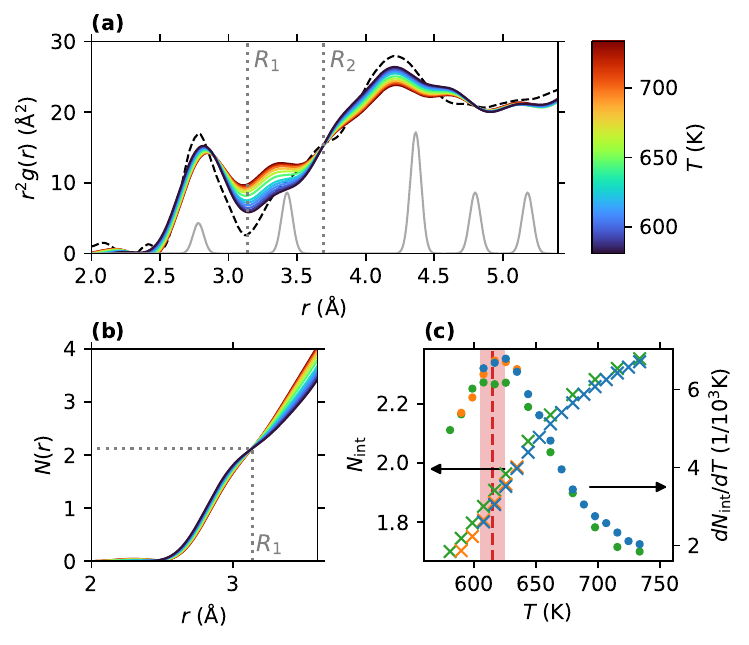}
    \caption{Temperature evolution of the structure of liquid Te. From dark red to dark blue the temperature decreases from 738 K to 585 K, as shown in the color bar. (a) shows the function $r^2g(r)$ and (b) shows the running coordination number $N(r)$. Also shown in (a) are trigonal Te~\cite{Adenis1989} (gray solid line) and electron diffraction results on amorphous Te~\cite{Ichikawa1973} (black dashed line).
    Vertical lines: we take $R_1$, the isosbestic point in (b), as the boundary of the first shell; $R_2$ is a near-isosbestic point in (a) and is taken to be the lower boundary of the second shell. (c) shows the number of atoms in the intermediate region (crosses, left $y$-axis) between $R_1$ and $R_2$ for all three scans, as well as its temperature derivative (dots, right $y$-axis). The derivative shows a peak close to the thermodynamic response maxima in Fig.~\ref{fig:Te_SAXS_summary} (dashed vertical line).}
    \label{fig:Te_coord_scan3}
\end{figure*}

Pertinent to the phenomenon of liquid polymorphism is the underlying structural transformation between the different liquid forms. The WAXS data, with examples given in Fig.~\ref{fig:Te_raw_scan1}a, allows us to examine the microscopic structural changes across the thermodynamic maxima in detail. Our experimental setup provides data up to $Q=7$\,\AA$^{-1}$, and previous works have shown that little change appears in $S(Q)$ beyond 6\,\AA$^{-1}$ across a range of 500 K~\cite{Akola2010,Endo2003}. Therefore, we append literature data from 7 to 15 \AA$^{-1}$~\cite{Akola2010,Menelle1987} to our results to obtain the radial distribution function $g(r)$ (further details in SI Appendix), which describes the average density at a distance $r$ from a reference atom. In Fig.~\ref{fig:Te_coord_scan3}a, we show the function $r^2 g(r)$ with cooling from 738 K to 585 K. Two salient features emerge: a decrease in the region between $\sim 3$ \AA\ and $\sim 3.7$ \AA, and an increase in the broad second peak from 3.7 to 4.7 \AA. In Fig.~\ref{fig:Te_coord_scan3}b we plot the running coordination number, $N(r)$, which is obtained from $g(r)$ and represents the averaged number of atoms within a distance $r$ from the reference atom. Indeed, the change in $N(r)$ is small up to $R_1 = 3.14$ \AA$^{-1}$, where the data appears to have an isosbestic point. We thus take $R_1$ to be the boundary of the first coordination shell, resulting in a coordination number of approximately 2.1 under all temperatures. Remarkably, the position of $R_1$ also lies close to the first minimum in $g(r)$ of amorphous Te~\cite{Ichikawa1973}, as shown in Fig.~\ref{fig:Te_coord_scan3}a, which further supports this choice of $R_1$ as the boundary of the first shell.

Here we note that earlier works attributed the structural transition in liquid Te to a decrease of coordination number from 3 to 2 upon cooling~\cite{Cabane1971}. However, this model has been much debated in the literature~\cite{Tsuzuki1995}, and it was pointed out that the measured coordination number depends sensitively on both the exact definition of the first shell boundary~\cite{Menelle1989} and the available $Q$ range~\cite{Hoyer1992}. In fact, more recent AIMD results~\cite{Akola2010} show that even the trend depends on the cutoff radius of the first shell: if the cutoff is at 3.1 \AA, the coordination number increases with cooling, while it decreases with a cutoff at 3.2 \AA. Notably, this is consistent with our observation that the coordination number remains constant with a cutoff around 3.14 \AA, leading to the conclusion that the coordination number alone does not explain the behavior of supercooled Te.

The plots shown in Figs.~\ref{fig:Te_coord_scan3}a and \ref{fig:Te_coord_scan3}b bear a notable resemblance to recent measurements on water~\cite{Benmore2019}. In particular, an isosbestic point was also found in the running coordination number $N(r)$ of water, leading to the conclusion that the first shell coordination is mostly constant from 300 K down to 244.2 K~\cite{Benmore2019}. In addition, a depletion of molecules was found upon supercooling in the region between the first and second peak, referred to as the interstitial region~\cite{Benmore2019}. This means that for both water and Te, intermediate-range ordering instead of changes in first-shell coordination drives the structural changes upon cooling and possibly distinguishes the two liquid structures: the high-temperature liquid is dominated by a denser and more disordered structure, including more interstitial atoms, while the low-temperature liquid is dominated by a less dense and more ordered structure up to the second shell.

To quantify the structural changes in liquid Te, we calculate the rate of depletion of atoms in between the first and second peaks, hereinafter referred to as the intermediate region. The lower boundary of this intermediate region is the first shell boundary $R_1$, while we define the upper cutoff $R_2$ as the near-isosbestic point observed in $r^2 g(r)$, around 3.69 \AA$^{-1}$. Although this choice of $R_2$ is somewhat arbitrary, small changes do not alter the conclusions below (see SI Appendix). 
We can then obtain the number of atoms between $R_1$ and $R_2$, $N_\mathrm{int}$, as a function of temperature. The results are plotted as crosses in Fig.~\ref{fig:Te_coord_scan3}c, and $N_\mathrm{int}$ increases with temperature as expected. Further, the temperature derivative of $N_\mathrm{int}$ can be obtained, as shown in the same panel, and it shows a peak similar in both position and shape to the maxima in Fig.~\ref{fig:Te_SAXS_summary}. This provides strong evidence that the thermodynamic changes in Fig.~\ref{fig:Te_SAXS_summary} are accompanied by a fast depletion of atoms in the intermediate region, signifying an increase in intermediate-range ordering.

\begin{figure*}
    \centering
    \includegraphics[width=0.7\textwidth]{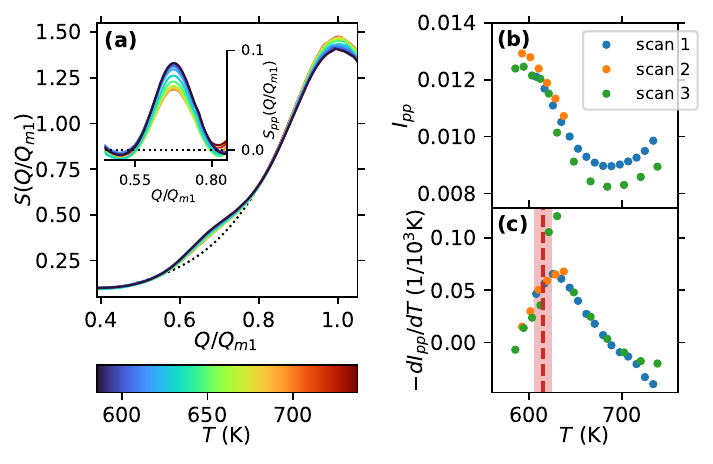}
    \caption{Analysis of the pre-peak. (a) The structure factor plotted as a function of reduced momentum transfer, $Q/Q_{m1}$, focusing on the low-$Q$ region. From dark red to dark blue the temperature decreases from 738 K to 585 K, as shown in the color bar. The dotted line shows the estimated background (see text). The inset shows the pre-peak region with the background subtracted. (b) The integrated pre-peak intensity, $I_{pp}$, and (c) the negative of its temperature derivative, both plotted as function of temperature. The derivative shows a peak close to the thermodynamic response maxima in Fig.~\ref{fig:Te_SAXS_summary} (dashed vertical line).}
    \label{fig:Te_prepeak}
\end{figure*}

The growing intermediate-range order can also be seen in the rise of the pre-peak shown in Fig.~\ref{fig:Te_raw_scan1}a. A closer look reveals that, with cooling, the center of the pre-peak moves to lower $Q$ similar to the first peak, $Q_{m1}$. Therefore, to compare the pre-peak at different temperatures, in Fig.~\ref{fig:Te_prepeak}a we plot the structure factor as a function of the reduced momentum transfer, $Q/Q_{m1}$. In order to better visualize the pre-peak, we subtract a background indicated by the dotted line. The background is modeled as a 3$^\text{rd}$-degree polynomial and fit with data before and after the pre-peak, $0.45 < Q/Q_{m1} < 0.55$ and $0.80 < Q/Q_{m1} < 0.90$, where $S(Q/Q_{m1})$ appears to change little with temperature (nevertheless, due to a possible temperature dependence, we specifically choose the data at $T \approx 610$ K for the fit). After background subtraction, the data in the pre-peak region, denoted here as $S_{pp}(Q/Q_{m1})$, is shown in the inset of Fig.~\ref{fig:Te_prepeak}a. Here, the near-Gaussian shape of the pre-peak and its growth with cooling becomes clear. Remarkably, the center position appears to remain the same at all temperatures. This means that the temperature dependence of the $Q$ position of the pre-peak follows almost exactly that of the pre-peak, suggesting that they may share a common origin in the local structure.

For a more quantitative analysis, we obtain the integrated pre-peak intensity, $I_{pp} = \int_{0.55}^{0.80} S_{pp}(q_r) d q_r$, where $q_r \equiv Q/Q_{m1}$. The temperature evolution of $I_{pp}$ is shown in Fig.~\ref{fig:Te_prepeak}b. Below approximately 690 K, $I_{pp}$ increases with cooling, gaining around 40\% towards the lowest temperature measured. The increase with heating above 690 K, on the other hand, may be due to temperature-dependent changes in the background. These change are difficult to model, and because the pre-peak is a small-amplitude feature overlaid on a rapidly rising background, as can be seen in Fig.~\ref{fig:Te_prepeak}a, performing temperature-dependent fitting of the background leads to noisy and model-dependent results. Therefore, we have chosen to subtract the same background for all temperatures. In Fig.~\ref{fig:Te_prepeak}c we plot the temperature derivative, $-dI_{pp}/dT$, where a peak can be seen close to the position of the thermodynamic response maxima. This signifies a rapid development of intermediate-range order consistent with the observations above. We note that, because a constant background is subtracted from all temperatures, the value of $-dI_{pp}/dT$ is independent from the modeling of this background. However, when interpreting the results, one should bear in mind that the temperature dependence of the actual background may contribute to the values of $-dI_{pp}/dT$; an accurate modeling of this background could be the subject of future work.

\section{Discussion and conclusions}
The results above have shown clear evidence for rapid intermediate-range ordering across the thermodynamic response maxima in supercooled Te, but details of this ordering remain to be discussed. Here we propose that bond-orientational order plays a crucial role. The pre-peak, centered around $Q_{pp} = 1.4$ \AA$^{-1}$, signals the presence of some structure on the order of $2\pi / Q_{pp} = 4.5$ \AA. In the crystal, this corresponds to the distance between next-nearest neighbors within a chain (i.e., the third peak of trigonal Te in Fig.~\ref{fig:Te_coord_scan3}a), whose bonds with their common neighbor form an angle of 103.2$^\circ$~\cite{Adenis1989}. The second peak in the liquid $g(r)$ is also close to 4.5 \AA, but slightly lower in $r$, indicating a somewhat smaller average bond angle consistent with AIMD results~\cite{Akola2010}.
With decreasing temperatures, the pre-peak grows as discussed above, 
while AIMD shows that the bond-angle distribution becomes sharply peaked around $96^\circ$~\cite{Akola2010}. In addition, the AIMD results have shown that two-coordinated atoms (within a soft cutoff around 3.2 \AA) have a bond angle distribution which is strongly peaked around 100$^{\circ}$, whereas atoms with higher coordination show a broader bond-angle distribution peaked close to 90$^{\circ}$~\cite{Akola2010}. This suggests that an increase in bond-orientational order would lead to a decrease in atoms in the intermediate region, as we have observed.

The similarities between Te and water in the various aspects we have discussed are worth emphasizing for several reasons. Firstly, the thermodynamic similarities indicate that the liquid phase diagrams of Te and water are likely similar. Thus, investigations into one may provide vital information for the other. Secondly, the similarities between the structural evolution of the two suggest a common microscopic origin. Our results confirm that a reduction in atoms in the intermediate region between the first and second peaks in $g(r)$, similar to the depletion of interstitial atoms observed in water~\cite{Benmore2019}, indeed accompanies the transition across the thermodynamic response maxima in liquid Te; such data is not yet available in water due to difficulties in reaching the deep supercooled state. Last but not least, the observed similarities strongly indicate that the anomalies of water are not limited to systems with hydrogen bonding or tetrahedral coordination, as is usually thought, but are general to a family of liquids. Finding the common features between Te and water may thus help generalize the origin of the observed phenomena. For example, it is well-known that water and other tetrahedral systems have an ``open'' local structure, while AIMD results have revealed an unusually large volume fraction of ``cavities'' in liquid Te as well~\cite{Akola2010}. On the other hand, the number of bonds is rather different between Te (2 to 3 per atom) and water (4 to 5 per molecule), and thus our observations indicate that the degree of thermodynamic anomalies cannot be predicted by bond number fluctuations alone.

Finally, we note that the enhanced intermediate-range ordering in the low-temperature limit is consistent with the two-state model mentioned above~\cite{Tanaka2012,Tanaka2020}, where the thermodynamic anomalies (such as the density maximum) are caused by an increasing amount of locally favored, albeit less dense, structures upon cooling. Since the presence of locally favored structures are expected in a number of liquids~\cite{Tanaka2020}, it is reasonable to predict that more systems can be found to exhibit water-type anomalies. Studying the commonalities and differences between these systems may help reveal the role of intermediate-range structures and bond-orientational ordering in the behavior of liquids in general.

In conclusion, using a combination of small-angle and wide-angle X-ray scattering, we have observed clear maxima in the thermodynamic response functions in supercooled liquid Te as well as the structural changes that accompany it. Our results suggest that intermediate-range ordering, likely of a bond-orientational nature, is strongly associated with the anomalous properties of liquid Te. This is consistent with the two-state model and points to the possibility of liquid polymorphism in this system. Moreover, the striking resemblance between Te and water in various properties, in spite of their different local structures, points to the possible existence of water-like thermodynamic anomalies in a family of liquid systems.
Thus, our study suggests that the competition between bond- and density-ordering can be a general phenomenon and plays a crucial role in the thermodynamics of liquid systems.

\section*{Materials and Methods}
The samples used in this study are Te powder (99.999\% metals basis, purchased from Alfa Aesar) vacuum-sealed in borosilicate glass ampoules. The inner diameter of the glass container is around 0.7 mm, and its wall thickness is around 0.085 mm. 

The experiment was performed at beamline 1-ID at the Advanced Photon Source. The X-ray energy was 76.112 keV. The planes of the WAXS and SAXS detectors are 2.003 m and 6.420 m from the sample, respectively. The $Q$-positions of the WAXS detector pixels are calibrated using a CeO$_2$ powder sample.

Further details on the experimental setup are provided in SI Appendix.

\section*{Acknowledgments}
We thank Charles L.\ Troxel Jr.\ for his tremendous help during preparation for the experiment. We also thank Ali Mashayekhi for his valuable support during the experiment.
This research used resources of the Advanced Photon Source, a U.S.\ Department of Energy (DOE) Office of Science User Facility at Argonne National Laboratory and is based on research supported by the U.S.\ DOE Office of Science-Basic Energy Sciences, under Contract No.~DE-AC02-06CH11357.
This work is supported by the U.S. Department of Energy, Office of Science, Office of Basic Energy Sciences under Contract No.~DE-AC02-76SF00515; the Deutsche Forschungsgemeinschaft (DFG, German Research Foundation) through the Collaborative Research Centre (CRC) 1242, project number 278162697; and the National Science Centre, Poland, grant agreement No.~2017/27/B/ST3/02860.

\bibliography{references}

\end{document}


\preprint{APS/123-QED}

\title{SI Appendix}

\date{\today}

\maketitle

\section{Experimental setup}
Figure~\ref{fig:setup} shows a schematics of the experimental setup used at beamline 1-ID at the Advanced Photon Source. The sample is placed in a hollow cylindrical oven wrapped in resistive heater wires. At the center of the tube, a through path is created for both the entrance and the exit of X-rays. A control thermocouple (TC) is inserted into a fixed location in the oven. Downstream from the sample is placed, in sequence, a WAXS detector (GE 41RT~\cite{Lee2008}), a helium flight tube, a beamstop (with a diode at the center to measure the transmitted beam intensity, $I_1$), and a SAXS detector (Pixirad2). Another diode (not shown), $I_0$, is placed before the sample and measures the incoming beam intensity. The planes of the WAXS and SAXS detectors are \SI{2.003}{m} and \SI{6.420}{m} from the sample, respectively. The $Q$-positions of the WAXS detector pixels are calibrated using a CeO$_2$ powder sample.

\section{Temperature calibration}
Due to the opening in the center of the oven for the passage of the X-rays, some temperature gradient is expected within the oven. In order to calibrate the sample temperature, we measure the melting of three materials, Bi, Te, and Sb, with the WAXS pattern. The temperature of the control TC is increased in small steps, and WAXS patterns are collected after equilibration at each step. Melting can be observed unambiguously as the disappearance of sharp Bragg peaks and the appearance of a halo pattern. The corresponding control TC temperature can be determined within \SI{5}{K}. In Fig.~\ref{fig:T_calib} we plot the temperature of the control TC thus determined against the melting temperature of each sample. The data are well described by the linear fit shown in the plot, and we use the fit result for the analysis. We note that the calibrated temperature range well covers the range used in our Te measurements, from above \SI{580}{K} to below \SI{750}{K}.

Another way to cross-check the temperature calibration and ensure consistency between different scans is through the position of the first peak of the liquid, $Q_{m1}$. This quantity depends sensitively on the sample temperature and can be determined with a high accuracy because of the relatively strong WAXS signal. In addition, it is independent of normalization and insensitive to possible smooth changes in the background.

In Fig.~\ref{fig:Te_Qm1}, we plot $Q_{m1}$ against the calibrated sample temperature. We can see that data from different scans align well with each other. From these data, the temperature derivative of $Q_{m1}$ is calculated and plotted in Fig.~3c in the main text.

\section{Density measurements}
The density of the sample is determined \emph{via} X-ray transmission measurements. Experimentally, we measure the incoming X-ray intensity $I_0$ and transmitted beam intensity $I_1$ at each sample temperature. The same measurements are also done with an empty capillary, giving the background values $I_0^\mathrm{bg}$ and $I_1^\mathrm{bg}$. We then have:
\begin{equation} \label{eq:transm}
    \frac{I_1}{I_1^\mathrm{bg}} = \frac{I_0}{I_0^\mathrm{bg}} \exp [-(\mu/\rho) \rho t],
\end{equation}
where $\mu/\rho$ is the mass attenuation coefficient of the sample, $\rho$ its density, and $t$ its thickness. For tellurium and an X-ray energy of \SI{76.112}{keV}, $\mu/\rho=\SI{3.68}{cm^2/g}$~\cite{Chantler1995a}.

With measured values of $I_0$, $I_1$, $I_0^\mathrm{bg}$, and $I_1^\mathrm{bg}$, Eq.~[\ref{eq:transm}] describes a relation between $\rho$ and $t$. Since $t$ depends on both the sample and the beam position, it may change between scans. Thus, for each cycle, $t$ is determined using the literature value of $\rho$ at the highest temperature~\cite{Tsuchiya1991}. The density $\rho$ at the other temperatures in the cycle can then be determined given this value of $t$. As can be seen in Fig.~\ref{fig:Te_scan1_cycles}c below and Fig.~2 in the main text, the results from different cycles and different scans are consistent with each other, and all agree well with the values reported by Tsuchiya~\cite{Tsuchiya1991}. This also indicates that sample movement is minimal during each cycle.

\section{Consistency between cycles}
As mentioned in the main text, each one of the three scans in our dataset consists of several cycles. In each cycle, the sample is first heated above its melting point and then cooled down in steps (usually \SI{10}{K}) to the lowest temperature. We take scan 1 as an example to demonstrate the consistency between data from different cycles, and the results are shown in Fig.~\ref{fig:Te_scan1_cycles}. Because of a high signal-to-noise ratio, the WAXS data from different cycles overlap well with each other, as shown in Fig.~\ref{fig:Te_scan1_cycles}a-b. The density measurements based on X-ray transmission are also well reproduced between cycles, as shown in Fig.~\ref{fig:Te_scan1_cycles}c.

In comparison, the SAXS signal is weaker and somewhat larger variations are observed between cycles, but the data remain consistent. As mentioned in the main text, in the analysis, the SAXS data at the highest temperature of each cycle (\SI{733}{K} in this scan) is subtracted from the data at other temperatures, and we analyze the difference
\begin{equation}
    \Delta S(Q) = S(Q) - S(Q; T=\SI{733}{K}).
\end{equation}
Figure~\ref{fig:Te_scan1_cycles}d shows $\Delta S(Q)$ integrated from $Q=\SI{0.06}{\per\angstrom}$ to \SI{0.15}{\per\angstrom}; the black crosses show the average values, with error bars representing the standard error of the mean. We then fit $\Delta S(Q)$ with a linear function:
\begin{equation}
    \Delta S(Q) = \Delta S(0) - k Q.
\end{equation}
The $Q=0$ intercept, $\Delta S(0)$, and the slope $k$ extracted from the fit are shown in Fig.~\ref{fig:Te_scan1_cycles}e and f, respectively. The same trend can be observed in panels (d), (e), and (f), where a maximum around \SI{615}{K} can be observed. The intercept $\Delta S(0)$ is reported in the main text and used for further analysis.

\section{WAXS analysis details}
This section shows details of the WAXS data analysis. To obtain $S(Q)$, we first calculate the azimuthally integrated intensity profile $I(Q)$ from the WAXS images. Next, we subtract the background measured with an empty glass ampoule without sample. The background consists of two parts: one part comes from the glass capillary and the air scattering after the sample, which is scaled by the X-ray transmission through the sample; the other part comes from air scattering before the sample which is not affected by the sample---because the sample is only about a millimeter in size while the air path is several meters long, the vast majority of the air scattering before the sample misses the sample in the WAXS geometry; only a negligible fraction, up to a few centimeters of air path immediately before the sample, adds a background affected by sample transmission. After background subtraction, we then normalize the curve, subtract the Compton scattering as well as a flat fluorescence background, and divide the data by the form factor $|f(Q)|^2$ to obtain the $S(Q)$ data. The resulting $S(Q)$ data for $T=\SI{724}{K}$ are shown in Fig.~\ref{fig:Te_WAXS_compare_723K}a as blue dots. Because a small amount of tellurium oxide is present, the data contain some weak powder rings which appear as sharp peaks in $S(Q)$. These outliers can be identified and excluded, resulting in the smoothed data shown as the red line, which agrees well with the X-ray data at \SI{723}{K} reported by Akola et al.~\cite{Akola2010}. The latter is also shown~\cite{Akola2010} to agree well with earlier neutron data by Menelle et al.~\cite{Menelle1989}. 

For our WAXS dataset, the measured $Q$-range is from around \SIrange{0.8}{7}{\per\angstrom}. However, it can be seen that~\cite{Akola2010}, across the wide temperature range from \SIrange{623}{1123}{K}, very little change occurs in $S(Q)$ above \SI{6}{\per\angstrom}. A similar observation can be made based on the data reported by Endo et al.~\cite{Endo2003} which included a smaller $Q$- and temperature-range. Therefore, in order to more accurately calculate the radial distribution function, $g(r)$, we append the previously reported data~\cite{Akola2010} of $S(Q)$ from \SIrange{7}{15}{\per\angstrom} to our results before integration. On the low-$Q$ side, a linear interpolation is used for the reduced structure function, $Q[S(Q)-1]$, between \SI{0}{\per\angstrom} and \SI{0.8}{\per\angstrom} before integration (see Fig.~\ref{fig:Te_WAXS_compare_723K}b). The result at $T=\SI{724}{K}$ can be compared with that at \SI{723}{K} reported by Akola et al.~\cite{Akola2010}, as shown in Fig.~\ref{fig:Te_WAXS_compare_723K}. A good agreement can be observed between the present data and previous results.

In Fig.~\ref{fig:Te_WAXS_compare_623K}, we show the same results at \SI{626}{K}, compared with the neutron diffraction data at \SI{623}{K} reported by Menelle et al.~\cite{Menelle1989,Akola2010}.

\section{Boundaries of the intermediate region}
In the main text, we have used $R_1=\SI{3.14}{\angstrom}$ and $R_2=\SI{3.69}{\angstrom}$ as the lower and upper boundaries of the intermediate region. Here, we show that the conclusions, in particular the rapid change in $N_\text{int}$ around the Widom line, do not depend sensitively on the definition of the boundaries. To test this, we change the value of $R_1$ or $R_2$ by \SI{\pm 0.2}{\angstrom}, and calculate the number of atoms for each pair of $R_1$ and $R_2$. The results are shown in Fig.~\ref{fig:Te_int_range_test}. Although different choices lead to different $N_\text{int}$ values (see Fig.~\ref{fig:Te_int_range_test}b), the derivatives all show a peak around the Widom line position (see Fig.~\ref{fig:Te_int_range_test}c).

\section{SAXS and WAXS data for all scans}
In Figs.~\ref{fig:Te_raw_scan2} and \ref{fig:Te_raw_scan3}, we show details of the WAXS and SAXS data for scans 2 and 3, respectively. The SAXS data in these scans show larger error bars because of lower signal levels, but the results are consistent with those from scan 1 shown in the main text.

The WAXS results for all three scans, including the calculated $g(r)$, are presented in Figs.~\ref{fig:Te_WAXS_scan1} to \ref{fig:Te_WAXS_scan3}. The results are consistent between different scans.

\begin{figure*}
    \centering
    \includegraphics{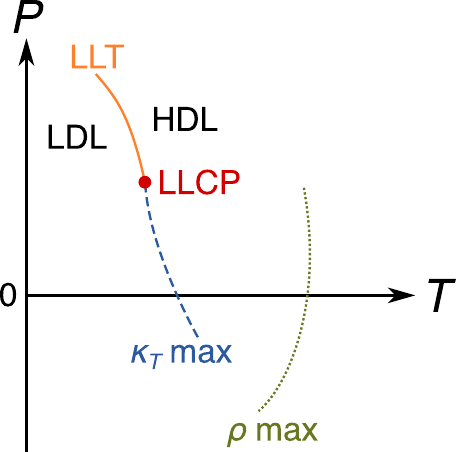}
    \caption{Schematics of a likely phase diagram of liquid water~\cite{Pallares2014,Gallo2016,Kim2017}. The line of liquid-liquid transition (LLT) separates the low-density (LDL) and high-density liquid (HDL) phases. The LLT ends at a liquid-liquid critical point (LLCP), with a critical pressure expected to be on the order of one to a few kilobars~\cite{Kim2017,Kim2020a}. From the LLCP emanates the line of isothermal compressibility maxima along isobars ($\kappa_T$ max), which intersects the $P \approx 0$ isobar around 229 K~\cite{Kim2017}. Also shown is the line of density maxima along isobars ($\rho$ max), which is known to intersect $P=1$ bar at $T=277$ K. Our study suggests that a similar phase diagram may be expected for liquid Te.}
    \label{fig:phase_diagram}
\end{figure*}

\begin{figure*}
    \centering
    \includegraphics[width=0.95\textwidth]{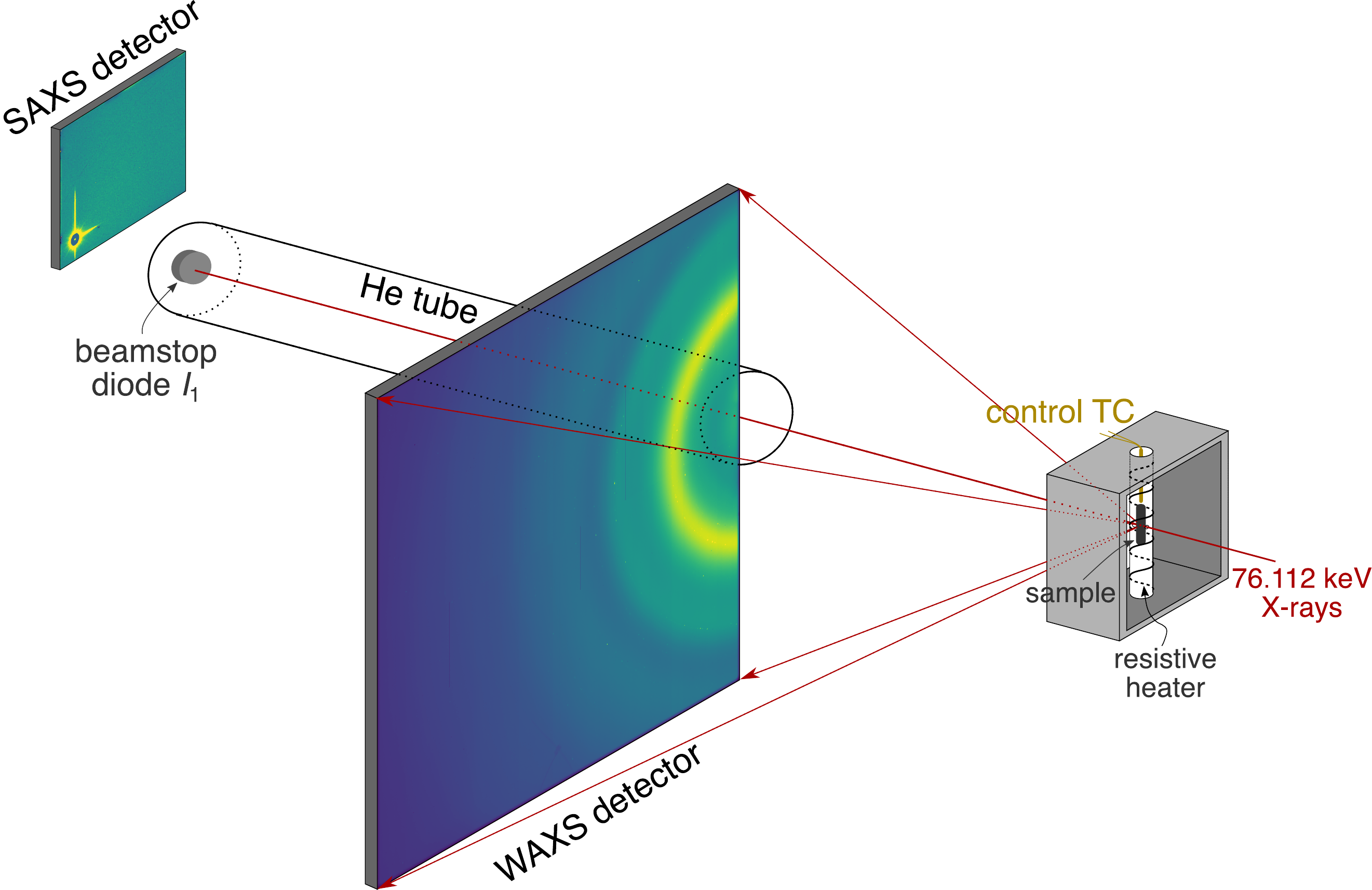}
    \caption{Experimental setup (not to scale).}
    \label{fig:setup}
\end{figure*}

\begin{figure*}
    \centering
    \includegraphics{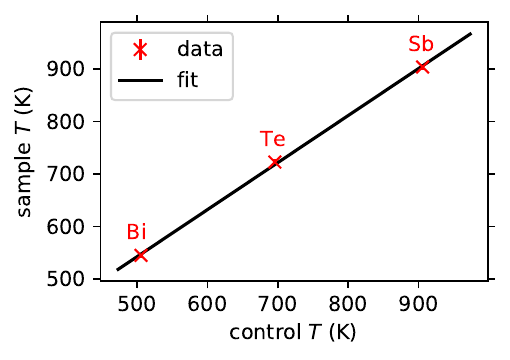}
    \caption{Temperature calibration. The red crosses show the temperature reading of the control TC at melting plotted against the actual melting temperature for each sample. The solid black line shows a linear fit.}
    \label{fig:T_calib}
\end{figure*}

\begin{figure*}
    \centering
    \includegraphics{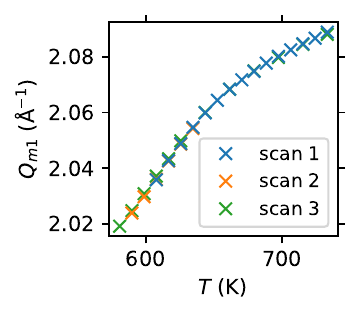}
    \caption{Position of the first peak, $Q_{m1}$, plotted against the calibrated sample temperature, $T$, for all three scans.}
    \label{fig:Te_Qm1}
\end{figure*}

\begin{figure*}
    \centering
    \includegraphics{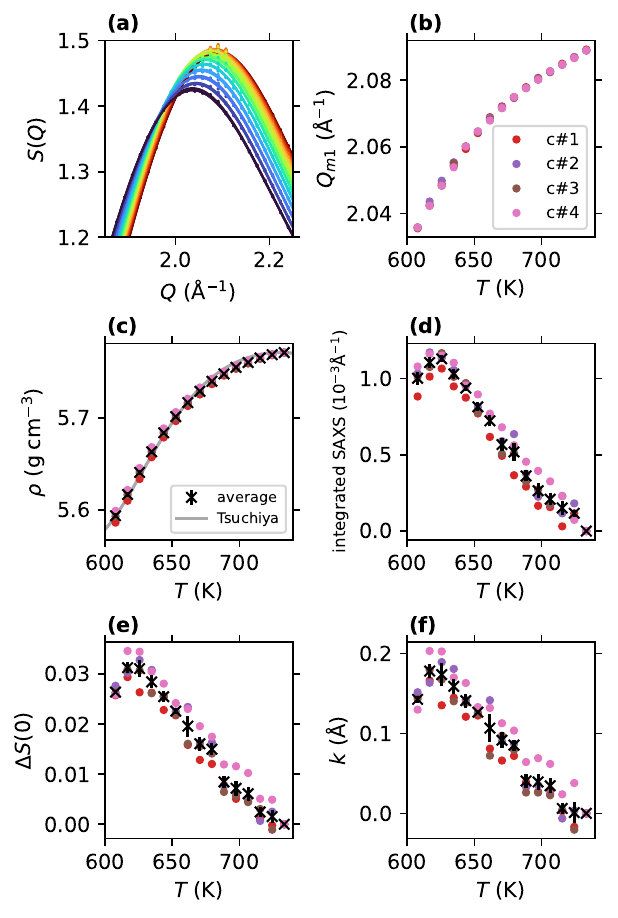}
    \caption{Consistency of data from different cycles in scan 1. (a) Dots: WAXS data around the first peak; lines: fit to the data with a Gaussian function and a linear background. From dark red to dark blue the temperature decreases from \SIrange{733}{608}{K}, same as Fig.~1 in the main text. Data and fits for all four cycles are shown, and the results well overlap. (b) The position of the first peak, $Q_{m1}$, extracted from the fits. The symbols representing different cycles overlap. (c) Density $\rho$ based on X-ray transmission measurements; also shown are the average (black crosses with error bars) as well as the results reported by Tsuchiya~\cite{Tsuchiya1991} (gray line). (d) SAXS data $\Delta S(Q)$ integrated from \SIrange{0.06}{0.15}{\per\angstrom}. (e) The intercept $\Delta S(0)$ and (f) the slope $k$ extracted from the linear fit to $\Delta S(Q)$.}
    \label{fig:Te_scan1_cycles}
\end{figure*}

\begin{figure*}
    \centering
    \includegraphics{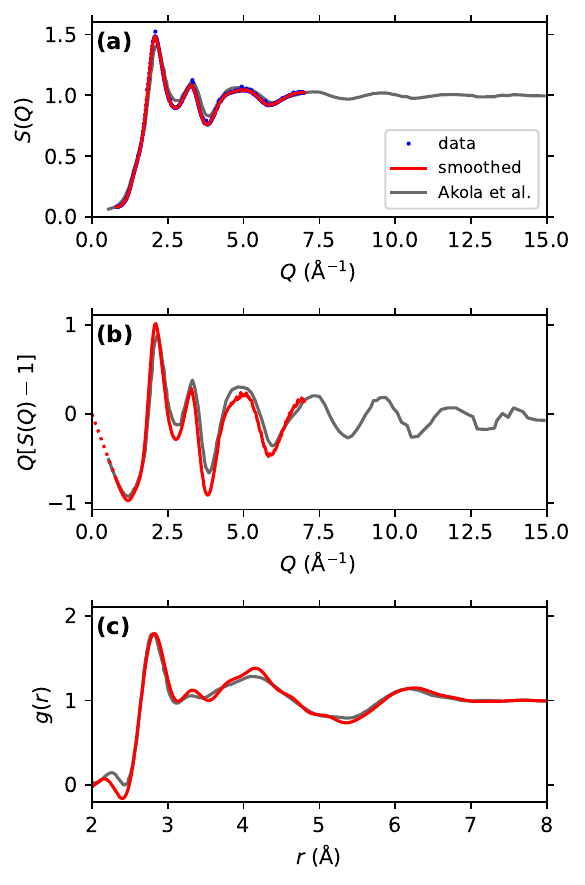}
    \caption{WAXS analysis. (a) Structure factor $S(Q)$. Blue dots: data at $T=\SI{724}{K}$. Red line: smoothed data after exclusion of sharp peaks. Gray line: X-ray diffraction data at \SI{723}{K} reported by Akola et al.~\cite{Akola2010}. (b) Reduced structure function, $Q[S(Q)-1]$. The dotted part shows the low-$Q$ extension used in the calculation of the radial distribution function, $g(r)$, shown in (c).}
    \label{fig:Te_WAXS_compare_723K}
\end{figure*}

\begin{figure*}
    \centering
    \includegraphics{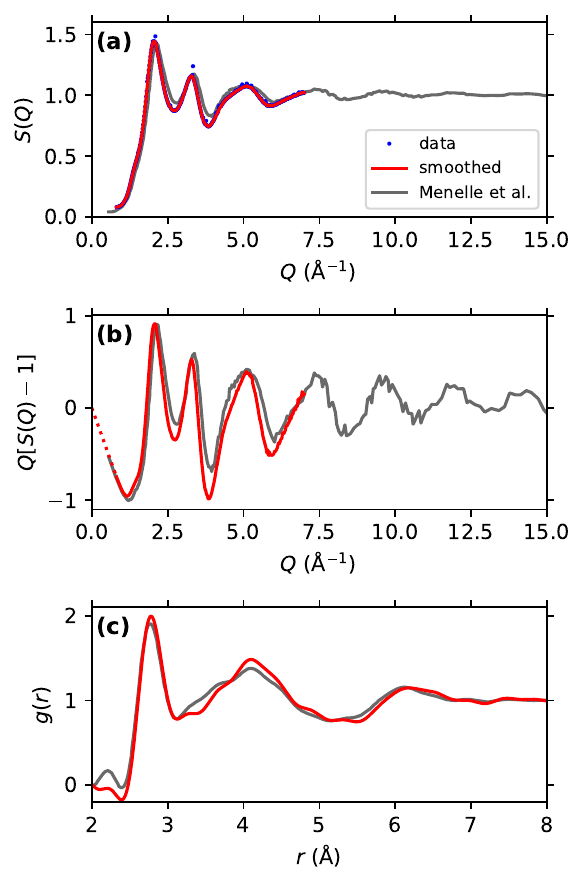}
    \caption{Same as Fig.~\ref{fig:Te_WAXS_compare_723K} but for $T=\SI{626}{K}$, compared with neutron diffraction data at \SI{623}{K} reported by Menelle et al.~\cite{Menelle1989,Akola2010}.}
    \label{fig:Te_WAXS_compare_623K}
\end{figure*}

\begin{figure*}
    \centering
    \includegraphics{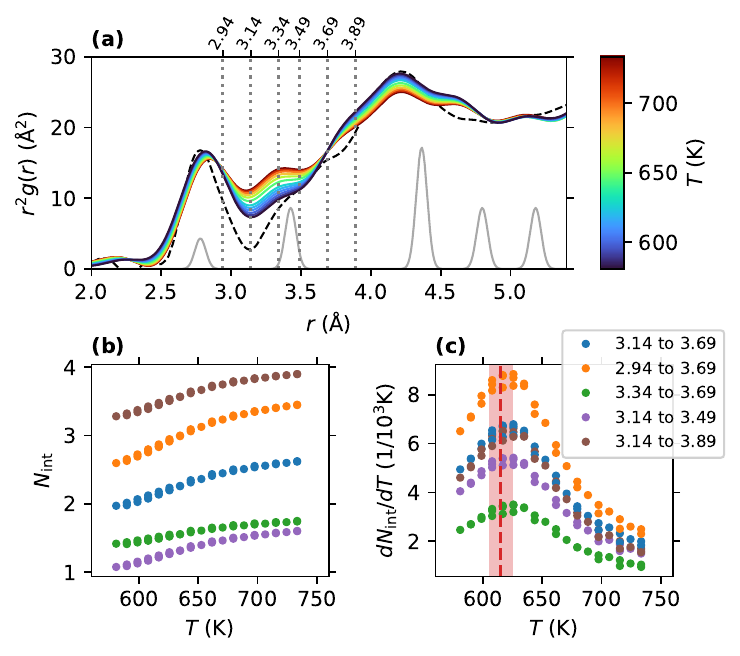}
    \caption{Test of the boundaries of the intermediate region. (a) The function $r^2 g(r)$. Colored solid lines: liquid Te from scan 3, from dark red to dark blue are decreasing temperatures from \SI{738}{K} to \SI{585}{K}, as shown in the color bar. Also shown are trigonal Te~\cite{Adenis1989} (gray solid line) and electron diffraction results on amorphous Te~\cite{Ichikawa1973} (black dashed line). Vertical dashed lines indicate different values of $R_1$ and $R_2$; the values are shown on the upper $x$-axis. (b) The number of atoms between $R_1$ and $R_2$, and (c) its temperature derivative. Different colors represent different choices of $R_1$ and $R_2$, as shown in the legend. Data from all three scans are included. Blue dots ($R_1=\SI{3.14}{\angstrom}$, $R_2=\SI{3.69}{\angstrom}$) correspond to the choice in the main text. The derivatives show a maximum near the Widom line (dashed vertical line), independent of the choice.}
    \label{fig:Te_int_range_test}
\end{figure*}

\begin{figure*}
    \centering
    \includegraphics{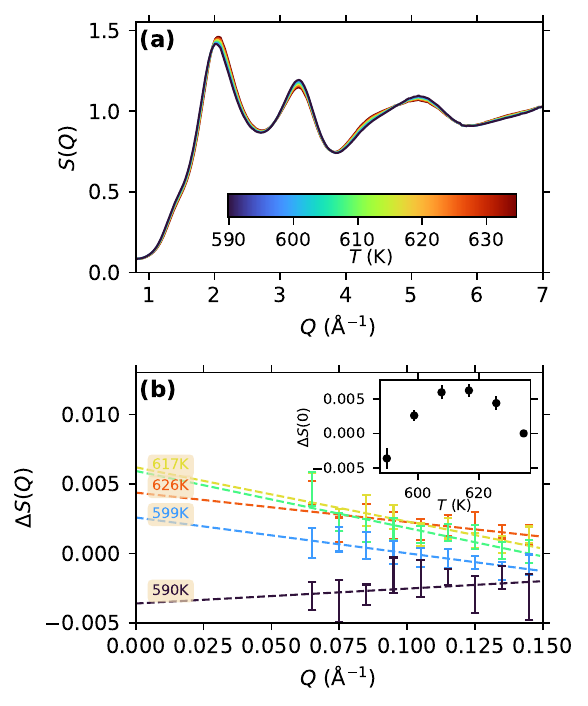}
    \caption{Same as Fig.~1 in the main text, but for scan 2. (a) Te WAXS profiles (structure factor $S(Q)$) from scan 2. From dark red to dark blue are decreasing temperatures as indicated by the color scale. (b) Te SAXS profiles from the same scan, subtracting the highest temperature profile (\SI{628}{K}) in the scan. The color scheme is the same as in (a). Data are shown with error bars, and the dashed lines show a linear fit to the data. Selected temperatures are marked next to the fit lines with the corresponding colors. Inset shows the intercept $\Delta S(0)$ plotted against sample temperature.}
    \label{fig:Te_raw_scan2}
\end{figure*}

\begin{figure*}
    \centering
    \includegraphics{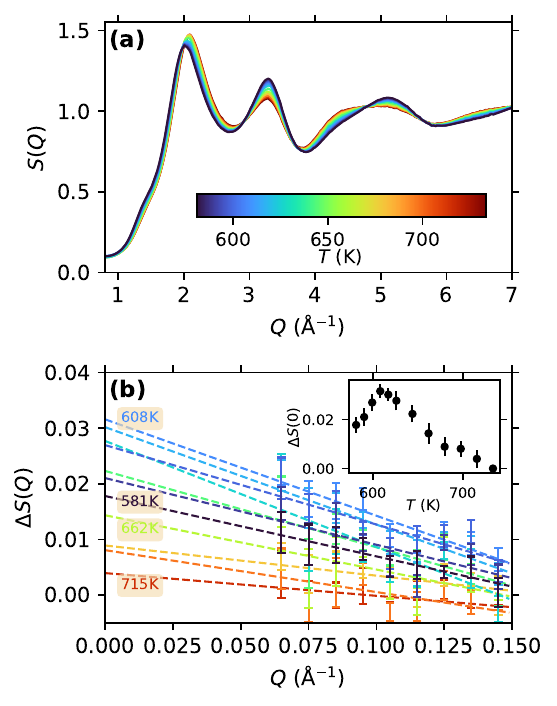}
    \caption{Same as Fig.~\ref{fig:Te_raw_scan2}, but for scan 3.}
    \label{fig:Te_raw_scan3}
\end{figure*}

\begin{figure*}
    \centering
    \includegraphics{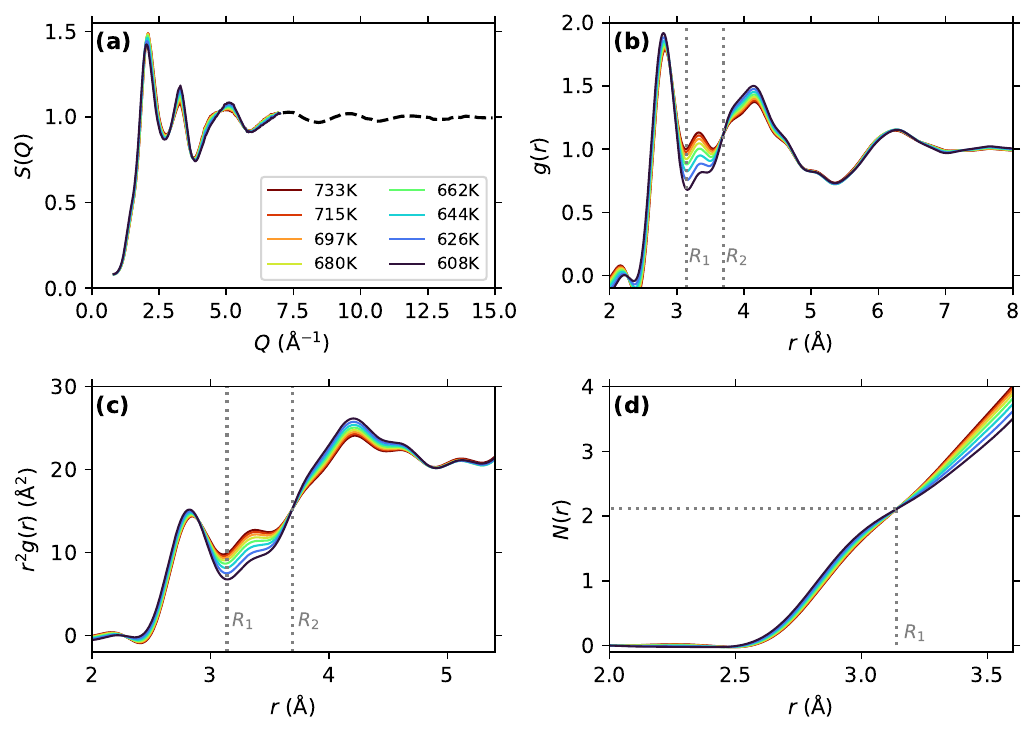}
    \caption{WAXS results for scan 1. (a) Structure factor $S(Q)$. The appended data~\cite{Akola2010} from \SIrange{7}{15}{\per\angstrom} is shown as dashed black line. (b) Radial distribution function $g(r)$. Dotted vertical lines shows the boundary of the intermediate region, $R_1$ and $R_2$, as discussed in the main text. (c) The function $r^2 g(r)$. (d) Running coordination number $N(r)$.}
    \label{fig:Te_WAXS_scan1}
\end{figure*}

\begin{figure*}
    \centering
    \includegraphics{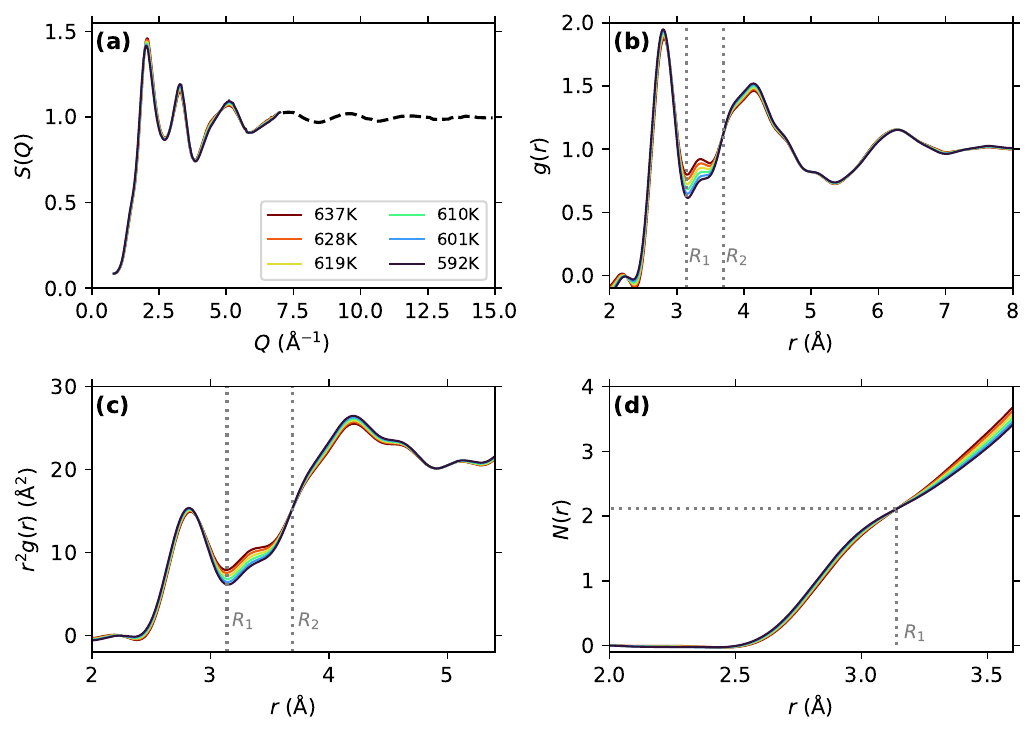}
    \caption{WAXS results for scan 2. Same as Fig.~\ref{fig:Te_WAXS_scan1}.}
    \label{fig:Te_WAXS_scan2}
\end{figure*}

\begin{figure*}
    \centering
    \includegraphics{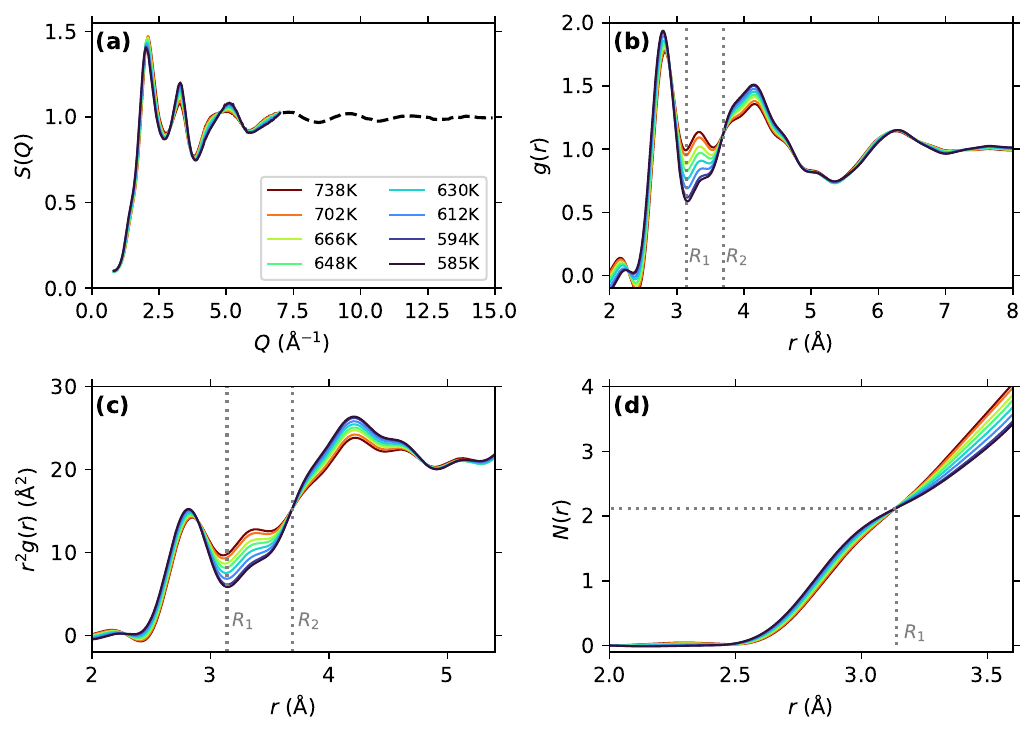}
    \caption{WAXS results for scan 3. Same as Fig.~\ref{fig:Te_WAXS_scan1}.}
    \label{fig:Te_WAXS_scan3}
\end{figure*}

\clearpage

\bibliography{references}